\documentclass[aps,twocolumn,showpacs,amsmath,amssymb]{revtex4-1}
\usepackage[T2A]{fontenc}
\usepackage{amsmath}
\usepackage{amssymb}
\usepackage{graphicx}
\usepackage{euscript}

\newlength{\pictwidth}
\begin{document}
\title{Recovery of ordered periodic orbits
with increasing wavelength
for sound propagation in a range-dependent waveguide}

\author{L.E. Kon'kov}
\author{D.V. Makarov}\email{makarov @ poi.dvo.ru}
\author{E.V. Sosedko}
\author{M.Yu. Uleysky}
\affiliation{
Laboratory of Nonlinear Dynamical Systems,\\
V.I.Il'ichev Pacific Oceanological Institute \\
of the Russian
Academy of Sciences, 690041 Vladivostok, Russia}

\begin{abstract}
We consider sound wave propagation in a range-periodic
acoustic waveguide in the deep ocean.
It is demonstrated that
vertical oscillations of a sound-speed perturbation,
induced by ocean internal waves,
influence near-axial rays in a resonant way, producing ray chaos
and forming a wide chaotic sea in the underlying phase space.
We study interplay between chaotic ray dynamics and wave motion
with signal frequencies of 50--100 Hz.
The Floquet modes of the waveguide are calculated
and visualized by means of the Husimi plots.
Despite of irregular phase space distribution of periodic orbits,
the Husimi plots
display the presence of ordered peaks within the chaotic sea.
These peaks, not being
supported by certain periodic orbits,
draw the specific ``chainlike'' pattern,
reminiscent of KAM resonance.
The link between the peaks and KAM resonance is confirmed
by ray calculations with lower amplitude of the sound-speed perturbation,
when the periodic orbits are well-ordered.
We associate occurrence of the peaks with
the recovery of ordered periodic orbits, corresponding to KAM resonance,
due to suppressing
of wavefield sensitivity to small-scale features of the sound-speed profile.
\end{abstract}
\maketitle

\section{\label{secIntro}Introduction}

In recent years interrelation
between wavefield structure and its semiclassical description
has attracted increasing attention in the context of
wave chaos --- wavefield manifestations of ray chaos.
Ray chaos means instability of ray trajectories,
conditioned by nonintegrability of the classical Hamiltonian equations
and is found to be the essential part of wave propagation
in various environments, ranging from artificial optical devices \cite{Wilc} to natural
media \cite{Klimes,Bott,Zas-ufn}.
In addition, the problem of wave chaos is closely connected
to the problem of quantum chaos, which is understood as quantum dynamics
of classically chaotic systems \cite{Shtockman}.

The present paper is devoted to long-range sound propagation in
the ocean, which have become of growing interest in recent decades
\cite{Chigarev,Palmer,Review,AET}.
Increasing of the sound speed with increasing of depth,
combined with the warming of the upper oceanic layer,
results in non-monotonic dependence of the sound speed on depth.
According to the Snell's law,
there occurs a waveguide confining acoustic waves
within a restricted water volume and preventing from their interaction
with the lossy bottom.
When we deal with guided wave propagation, weak
inhomogeneities along the axis of propagation may be sufficient for dividing
phase space into regular and irregular regions \cite{Viro2001}.
This division, relying on the KAM theory,
even persists in the case of a stochastically-perturbed waveguide \cite{PRE,JPA},
giving rise for coherent ray clusters \cite{Chaos}.
Ray chaoticity leads to
smearing of a spatial wavefield structure due to
irregular mode coupling \cite{ViroPRE99,ViroPRE2005},
random-like distribution
of ray arrival times at the receiver \cite{Viro-times,Brown},
or anomalous transmission loss due to the
chaos-assisted ray escaping from a waveguide \cite{Vadov,Acoust}.

On another front, it is well established that
a wavepacket may demonstrate coherent or incoherent
motion depending upon whether the initial position of a wavepacket is in the
regular or irregular part of classical phase space \cite{Latka,BZ}.
The packet initially concentrated inside a region of stability remains localized,
while the packet placed within a chaotic region spreads rapidly over all the chaotic layer.
In contrast to the semiclassical limit, boundaries between stable and chaotic regions are
penetrable at nonzero wavelength. This enables
extension of a wavepacket, evolving in the chaotic region, into
area with regular dynamics;
the effect is amplifying with increasing wavelength \cite{Backer}.
Thus wave corrections imply suppressing
of the phase space separation.

Chaos means irregular behavior of rays and can be thought of
as purely refractional phenomena.
Description of chaos-induced effects in wave motion far from the semiclassical limit
requires the understanding of how wave refraction
depends on wavelength.
This issue is of great importance
in the presence of small-scale features, which seem to be
irrelevant for wave refraction at low frequencies \cite{Hege}.
In underwater acoustics, these features are usually associated
with internal waves.
In the present paper we follow two aims.
First, we study the effect
of small-scale vertical oscillations of a perturbation on ray dynamics.
We shall show that these oscillations account
for strong chaos of near-axial rays.
Second, we investigate interrelation between strong chaos
of near-axial rays and
the wavefield structure at low frequencies.

The paper is organized as follows. In the next section we
describe briefly the model of a waveguide.
Section III is devoted to classical ray dynamics.
In Section IV we study wavefield properties by means
of Husimi representation of the Floquet modes.
In Conclusion we shortly discuss the results obtained.

\section{The model of a waveguide}

Consider a monochromatic wave-field in a two-dimensional acoustic waveguide
in the deep ocean
with the sound speed $c$ presented in the form
\begin{equation}
c(z,\,r)=c_0+\Delta c(z)+\delta c(z,\,r),
\label{c}
\end{equation}
where $c_0$ is a reference sound speed, $\Delta c(z)$ represents the range-independent depth change
of the sound speed due to the waveguide, and $\delta c(z,\,r)$ is a small term varying with range $r$.
In the present paper we give consideration to the narrow--angle wave propagation,
when the original Helmholtz equation for a wavefield reduces to the parabolic equation
\begin{equation}
\begin{aligned}
&\frac{i}{k_0}\frac{\partial\phi(z,\,r)}{\partial r}=\hat H\phi(z,\,r),
\\
&\hat H=-\frac{1}{2k_0^2}\frac{\partial^2}{\partial z^2}+\frac{\Delta c(z)+\delta c(z,\,r)}{c_0},
\end{aligned}
\label{parabolic}
\end{equation}
Here $k_0=2\pi f/c_0$ is the wavenumber
in the reference medium with $c=c_0$, $f$ is signal frequency.
The parabolic equation formally coincides with
the non-stationary Shr\"odinger equation.
In this analogy one
treats range $r$ as the time-like variable, $\Delta c(z)$
as an unperturbed potential,
$\delta c$ as a time-dependent perturbation, and $k_0^{-1}$
as the Planck constant.

\begin{figure}[!t]
\begin{center}
\centerline{
\includegraphics[width=0.24\textwidth]{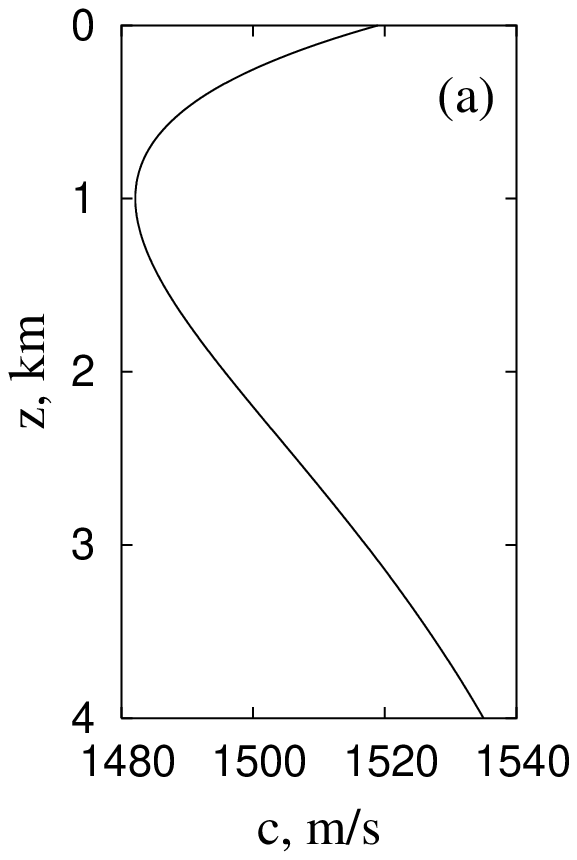}
\includegraphics[width=0.24\textwidth]{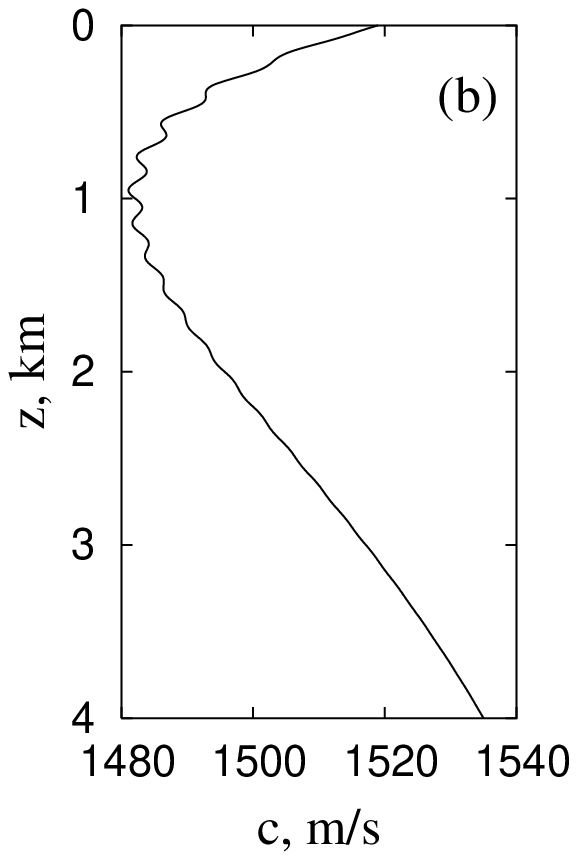}
}
\caption{Sound-speed profiles:
a) at $r=0$,
b) at $r=1.25$~km.
}
\label{profs}
\end{center}
\end{figure}

In the present paper we shall consider an idealistic
model of a waveguide with
$\Delta c(z)$ and $\delta c(z,\,r)$
given by the following expressions
:
\begin{equation}
\Delta c(z)=
-\dfrac{c_0b^2}{2}(\mu-e^{-az})(e^{-az}-\gamma),
\label{prof}
\end{equation}
\begin{equation}
\delta c(z,\,r)=\varepsilon c_0\frac{z}{B}\,e^{-2z/B}
\sin\frac{2\pi z}{\lambda_z}\sin\frac{2\pi r}{\lambda_r},
\label{dc}
\end{equation}
where  $c_0=c(z=h)=1535$~m/s,
$\gamma =\exp(-ah)$, $h=4.0$~km is depth of the ocean bottom, $\mu=1.078$,
$a=0.5$~km$^{-1}$, $b=0.557$,
$\varepsilon=0.005$, $B=1$~km, $\lambda_z=0.2$~km, $\lambda_r=5$~km.
Function $\Delta c(z)$ takes on the smallest value at the depth
\begin{equation}
z_a=\frac{1}{a}\ln\frac{2}{\mu+\gamma}\simeq1 \text{~km}.
\label{za}
\end{equation}
We shall refer this depth as the channel axis.
The respective unperturbed sound-speed profile is depicted
in Fig.~1(a).
Fast oscillations of $\delta c(z,\,r)$
are included in order to mimic the effect of
internal wave fine structure and distort the sound-speed profile,
as it is demonstrated in Fig.~1(b).

\section{Ray dynamics}\label{rays}

The classical counterpart of the operator $\hat H$ is the Hamiltonian
\begin{equation}
H=-1+\frac{p^2}{2}+\frac{\Delta c(z)}{c_0}+\frac{\delta c(z,\,r)}{c_0},
\label{h}
\end{equation}
where $p=\tan\alpha$ is the analog to mechanical momentum, $\alpha$ is a grazing angle of a sound ray.
Ray trajectories obey the Hamiltonian equations
\begin{equation}
\dfrac{dz}{dr}=\dfrac{\partial H}{\partial p}=p,
\label{sys1}
\end{equation}
\begin{equation}
\dfrac{dp}{dr}=-\frac{\partial H}{\partial z}=-\frac{1}{c_0}\frac{d\Delta c}{dz}-\frac{1}{c_0}\frac{d\delta c}{dz}.
\label{sys2}
\end{equation}
The last term in the right-hand side of (\ref{sys2}) can be rewritten in the following form:
\begin{equation}
\begin{aligned}
\frac{1}{c_0}\frac{d\delta c}{dz}=\frac{\varepsilon e^{-2z/B}}{2B}\biggl[\biggr.
&\left(1-\frac{2z}{B}\right)\left(\cos\psi^--\cos\psi^+\right)-
\\
&-k_zz\left(\sin\psi^--\sin\psi^+\right)
\biggl.\biggr],
\end{aligned}
\label{sys22}
\end{equation}
where we denoted $\psi^\pm=k_zz\pm k_rr$, $k_z=2\pi/\lambda_z$, $k_r=2\pi/\lambda_r$.
The smallness of $\lambda_z$ implies that
the range-dependent term oscillates rapidly
along a ray path, except for the resonant regions, where
either the condition
\begin{equation}
\frac{d\psi^+}{dr}=k_zp+k_r\simeq0
\label{res}
\end{equation}
or the condition
\begin{equation}
\frac{d\psi^-}{dr}=k_zp-k_r\simeq0,
\label{res-m}
\end{equation}
is fulfilled.
The theory of such resonances
was developed in \cite{Acoust,Itin,Neishtadt,JETPL,Mezic,PRE07,CNS}.
Here we only give a brief description of their properties.

\begin{figure}[!t]
\begin{center}
\centerline{
\includegraphics[width=0.43\textwidth]{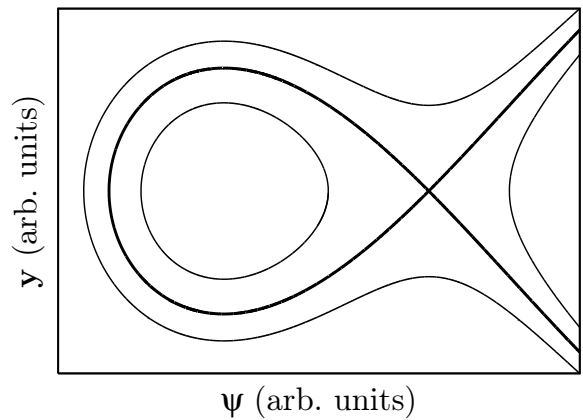}}
\caption{The phase portrait corresponding to
the Hamiltonian (\ref{h-tilde})
}
\label{pendulum}
\end{center}
\end{figure}

Let us consider one of the resonant conditions,
for instance, the former one.
First we simplify the equation (\ref{sys22}). Since $k_z$ can be thought of as a large parameter,
we leave only those terms
in the right-hand side, which are proportional to $k_z$. In addition,
we neglect the non-resonant term $\sim\sin\psi^-$.
Thus we obtain
\begin{equation}
\frac{dp}{dr}=
-\frac{1}{c_0}\frac{d\Delta c}{dz}
-\frac{\varepsilon k_zze^{-2z/B}}{2B}\sin\psi^+.
\label{p-cut}
\end{equation}
At the next step we shall describe variations of the perturbation phase
$\psi^+$ along a ray path.
Using (\ref{sys22}) and (\ref{res})
and omitting superscript ``+'',
one derives the pendulum-like equation
\begin{equation}
\frac{d^2\psi}{dr^2}+\frac{\varepsilon k_z^2ze^{-2z/B}}{2B}\sin\psi
+\frac{k_z}{c_0}\frac{d\Delta c}{dz}=0.
\label{pend-like}
\end{equation}
This equation can be rewritten as the coupled pair of first-order
equations
\begin{equation}
\begin{aligned}
\frac{d\psi}{dr}&=y,\\
\frac{dy}{dr}&=-\frac{\varepsilon k_z^2ze^{-2z/B}}{2B}\sin\psi
-\frac{k_z}{c_0}\frac{d\Delta c}{dz},
\end{aligned}
\label{sys-t}
\end{equation}
corresponding to the Hamiltonian $\tilde H$ of the form
\begin{equation}
\tilde H(y,\,\psi)=\frac{y^2}{2}+\frac{k_z}{c_0}\frac{d\Delta c}{dz}\psi
-\frac{\varepsilon k_z^2ze^{-2z/B}}{2B}\cos\psi,
\label{h-tilde}
\end{equation}
where $y$ and $\psi$ are treated as canonically conjugated
momentum and coordinate, respectively.
If the inequality
\begin{equation}
\left|\frac{1}{c_0}\frac{d\Delta c}{dz}\right|<\frac{\varepsilon k_zze^{-2z/B}}{2B}
\label{ineq}
\end{equation}
is satisfied, then the phase portrait corresponding to the Hamiltonian
(\ref{h-tilde})
contains a resonant area bounded by the separatrix loop (see Fig.~\ref{pendulum}).
A ray can cross the separatrix due to variation of depth $z$,
included in Eqs. (\ref{pend-like})--(\ref{ineq}) as a slowly-varying parameter.
When a ray arrives the resonant area, $\psi$ switches
its behavior from rotation
to oscillation, that is followed by localization of ray momentum
in a narrow interval near the resonant value $p_\text{res}=-k_r/k_z$.
The emphatic point is that each crossing of the resonant area
is followed by a jump-like variation of the ray Hamiltonian (\ref{h}),
which depends extremely on the initial conditions;
therefore multiple visits to the resonant area cause chaotic ray diffusion in the
underlying phase space.

\begin{figure}[!t]
\begin{center}
\centerline{
\includegraphics[width=0.45\textwidth]{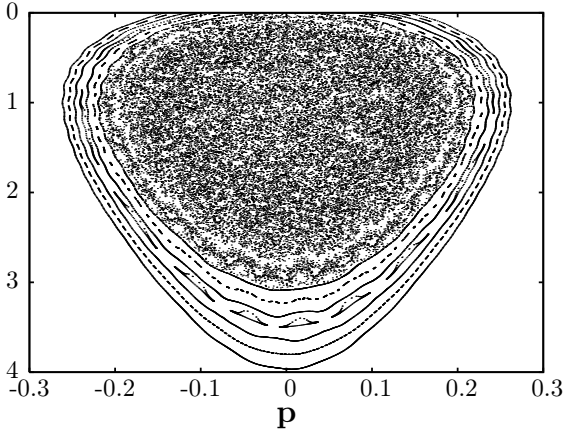}}
\caption{Poincar\'e map with $\varepsilon=0.005$.}
\label{poinc}
\end{center}
\end{figure}

The formulae (\ref{res}) and (\ref{ineq}) allow one to distinguish
the rays affected by the resonance.
The inequality (\ref{ineq}) is fulfilled only near the waveguide axis,
where $d\Delta c/dz=0$.
The ray momentum at the axis is given by the equation
\begin{equation}
p(z=z_a,\,H_0)=\sqrt{2E+\frac{b^2(\mu-\gamma)^2}{4}},
\label{p_ax}
\end{equation}
where $z_a$ is the depth of the channel axis (\ref{za}), and
the parameter
\begin{equation}
E=1+H\simeq1+H|_{\delta c=0}
\label{Edef}
\end{equation}
can be referred to as the ``energy'' of ray oscillations in a waveguide.
Substituting (\ref{p_ax}) into (\ref{res}), we find
the resonant value of $E$
\begin{equation}
E_\text{res}=\frac{\lambda_z^2}{2\lambda_r^2}-\frac{b^2(\mu-\gamma)^2}{8}.
\label{res-h}
\end{equation}
Note that formula (\ref{res-h}) relates to the resonance (\ref{res-m}) as well.
The conditions (\ref{res}) and (\ref{res-m})
differ by the sign of resonant momentum, the former arises at $p<0$,
while the latter takes place at $p>0$.
It means that these resonances
affect the same rays, but they act at different phases of a trajectory.
With $\lambda_z=0.2$~km and $\lambda_r=5$~km the condition (\ref{res-h})
holds for near-axial rays.
According to Fig.~\ref{poinc}, resonances
(\ref{res}) and (\ref{res-m}) cause wide chaotic sea
in phase space, without considerable stable islands within.
Therefore one can assume that
chaotic diffusion of rays inside the sea is close to ergodic mixing.

\section{Floquet modes}

Mixing ray dynamics
anticipates fast decoherence and spreading of a wavepacket initially
located within the chaotic sea,
until all the area of the sea will be covered \cite{Latka}.
Certainly, it is the case in the short wavelength limit,
when ray and wave descriptions are well correlated \cite{BZ,Prants}.
The question we ask is
how the chaotic sea reveals
itself at relatively low
frequencies, when influence of
diffraction and interference is non-negligible,
and one to one correspondence
between a wave pattern and its semiclassical approximation
shouldn't be expected.
To address this issue, we shall analyze
phase space structure of the Floquet modes,
which were first applied for studying underwater sound propagation
in \cite{ViroChaos2004,ViroPRE2005}.
The Floquet modes can be cast in the form:
\begin{equation}
u_m(z,r)=e^{i\alpha_mr/\lambda_r}\Psi_m(z,r),
\label{fl}
\end{equation}
where $m=1,2,...$, $\Psi_m(z,r)=\Psi_m(z,r+\lambda_r)$,
$\alpha_m$ is a real constant.
The Floquet modes $u_m$ are the eigenfunctions of the shift operator $\hat F$,
defined as
\begin{equation}
\hat F\phi(z,r)=\phi(z,r+\lambda_r).
\label{shift}
\end{equation}
In the present paper we
consider the functions
\begin{equation}
\Phi_m(z)=\Psi_m(z,r=0).
\label{fl-0}
\end{equation}
Each of these functions can be expanded
in some orthogonal basis
\begin{equation}
\Phi_m(z)=\sum_l c_{lm}\phi_l(z),
\label{fl-modexp}
\end{equation}
where $c_{1m},c_{2m},...$ are the components of
$m$-th eigenvector of the matrix with elements
\begin{equation}
F_{mn}=\int\limits_{z=0}^{h}\,\phi_m(z)\hat F\phi_n(z)\,dz.
\label{fl-matr}
\end{equation}
Here $\hat F\phi_n(z)$ is the solution of the parabolic equation
at the range $r=\lambda_r$ with initial condition
$\phi(z,r=0)=\phi_n(z)$.
We found the set of functions $\phi_n$ by solving
the Sturm-Liouville problem
\begin{equation}
\frac{\partial^2 \phi_n}{\partial z^2}+
2k_0^2\left(E_n-\frac{\Delta c}{c_0}\right)\phi_n=0.
\label{St-L}
\end{equation}
The solution is the following:
\begin{equation}
\phi_n(\xi)=A_ne^{-\xi/2}\xi^{s_n}G(1-n,\,2s_n+1,\,\xi),
\label{nmodes}
\end{equation}
where $n$ is a positive integer,
$\xi$ is linked with depth by formula
\begin{equation}
\xi(z)=\frac{2k_0}b{a}e^{-az},
\label{xi}
\end{equation}
$A_n$ is the normalization constant,
determined by the condition
\begin{equation}
\int \phi_n\phi_n^*\,dz=1,
\label{norm}
\end{equation}
$G(1-n,\,2s_n+1,\,\xi)$ is the degenerate
hypergeometric function,
and the parameter $s_n$ is given by the expression
\begin{equation}
s_n=\frac{k_0}{a}\sqrt{\frac{\mu\gamma b^2}{2}-E_n}.
\label{sn}
\end{equation}
Eigenvalues of the parameter $E$ are given by the following expression
\begin{equation}
E_n=\frac{\mu\gamma b^2}{2}-\frac{1}{2}\left[b\frac{\mu+\gamma}{2}-\frac{a}{k_0}
\left(n+\frac{1}{2}\right)\right]^2.
\label{en}
\end{equation}
At small $n$ the functions $\phi_n$ coincide with normal modes
of the unperturbed waveguide.

Phase space representation of the Floquet modes can be obtained by use of
the Husimi distribution function
\begin{equation}
\begin{aligned}
W_h(z,\,p,\,r)=&\Biggl|\Biggr.\frac{1}{\sqrt[4]{2\pi\Delta_z^2}}\int\,dz'\phi_n(z',\,r)
\\
&\exp\left(ik_0p(z'-z)-\frac{(z'-z)^2}{4\Delta_z^2}\right)
\Biggl|\Biggr.^2.
\end{aligned}
\label{Husimi}
\end{equation}
Here $\Delta_z$ is the smoothing scale, which we took of 100 m.

Before analyzing the Floquet modes,
it is necessary to define a criterion
by which we shall determine
their chaoticity.
The simplest way is to compare phase space location
of a Floquet mode with classical phase space structures \cite{Ketz}).
We shall use
the criterion of Leboeuf and Voros \cite{LebVor,Arranz,Biswas},
identifying chaotic Floquet state 
by irregular distribution of zeroes of the Husimi function.
Husimi zeroes of regular Floquet states are located along curves.
For the sake of a convenient representation, we shall analyze distribution
of Husimi zeroes in the space of the action and angle variables.
The action and angle variables are introduced by the following
formulae
\begin{equation}
I=\frac{1}{2\pi}\oint p\,dz,\quad
\theta=\frac{\partial}{\partial I}\int\limits_{z_0}^z p\,dz.
\label{acangle}
\end{equation}
The action variable $I$ measures steepness of a ray trajectory
and is equal to 0 for the horizontal axial ray.
In the integrable limit all the Husimi zeroes are distributed
along horizontal lines $I=\text{const}$.
We shall restrict ourselves by only qualitative analysis of the Husimi
zeroes in the range of small values of the action, where
the chaotic sea takes place.
The exact analytical expressions for the action and angle variables
are presented in Appendix.

In addition, we shall calculate the so-called number of principal components
\cite{Sugita}.
For the Floquet mode with number $m$, it is determined by the formula
\begin{equation}
\Gamma(m)=\frac{1}{\sum\limits_l |c_{lm}|^4}.
\label{npc}
\end{equation}
This quantity measures delocalization of a Floquet mode with respect
to the basis of the eigenfunctions (\ref{nmodes}), and also may be treated
as a measure of chaoticity. Roughly speaking, $\Gamma$
is the number of eigenmodes of the unperturbed waveguide, which give the dominant
contribution into a given Floquet mode.
Chaotic Floquet modes, being formed by large number of waveguide's
eigenmodes, are characterized
by large values of $\Gamma$ \cite{ViroPRE2005}.

\begin{figure}[!t]
\begin{center}
\centerline{
\includegraphics[width=0.45\textwidth]{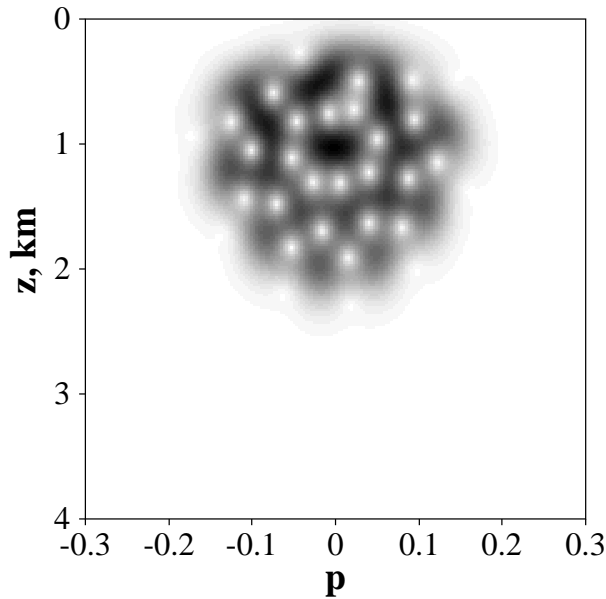}}
\caption{Floquet mode the largest number of principal components.}
\label{unst}
\end{center}
\end{figure}

\begin{figure}[!t]
\begin{center}
\includegraphics[width=0.45\textwidth]{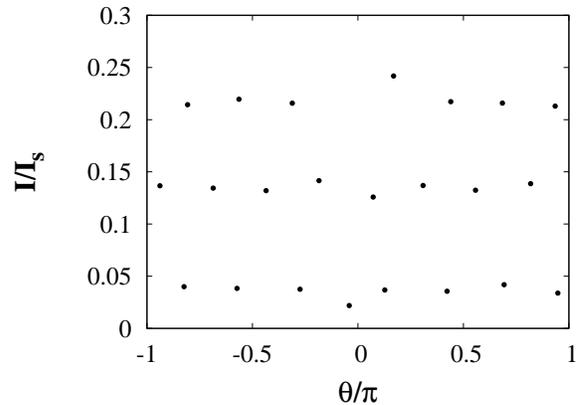}
\caption{Distribution of Husimi zeroes for the Floquet mode depicted in
Fig.~\ref{unst} in the plane
of normalized initial values of the action and angle.
$I_s$ is the most accessible value of the action for guided rays.
}
\label{zero1}
\end{center}
\end{figure}

Let us start with considering the frequency of 100~Hz.
Despite of strong chaos in the ray limit, the majority
of the Floquet modes are localized within narrow bands 
in phase space with $\Gamma$ ranging from 1 to 2.
Only few of Floquet modes spread over 
the chaotic layer.
The most extended Floquet mode, having the largest
number of principal components ($\Gamma\simeq4$),
is presented in Fig.~\ref{unst}.
It's Husimi zeroes, though covering wide area,
are located along almost horizontal curves, that is shown in Fig.~\ref{zero1}.
Since that we can regard this mode rather as weakly irregular than chaotic.

\begin{figure}[!t]
\begin{center}
\includegraphics[width=0.45\textwidth]{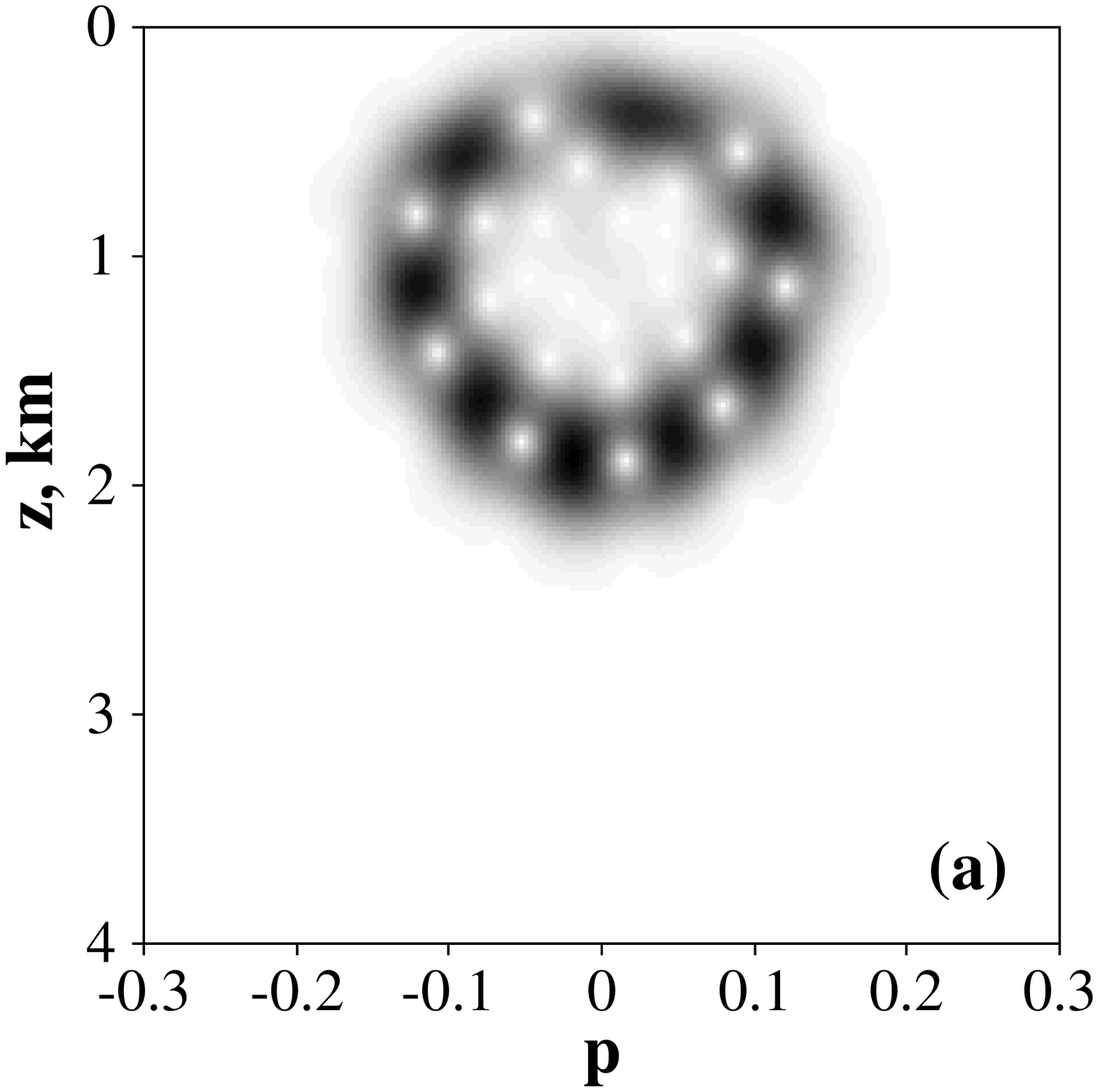}
\includegraphics[width=0.45\textwidth]{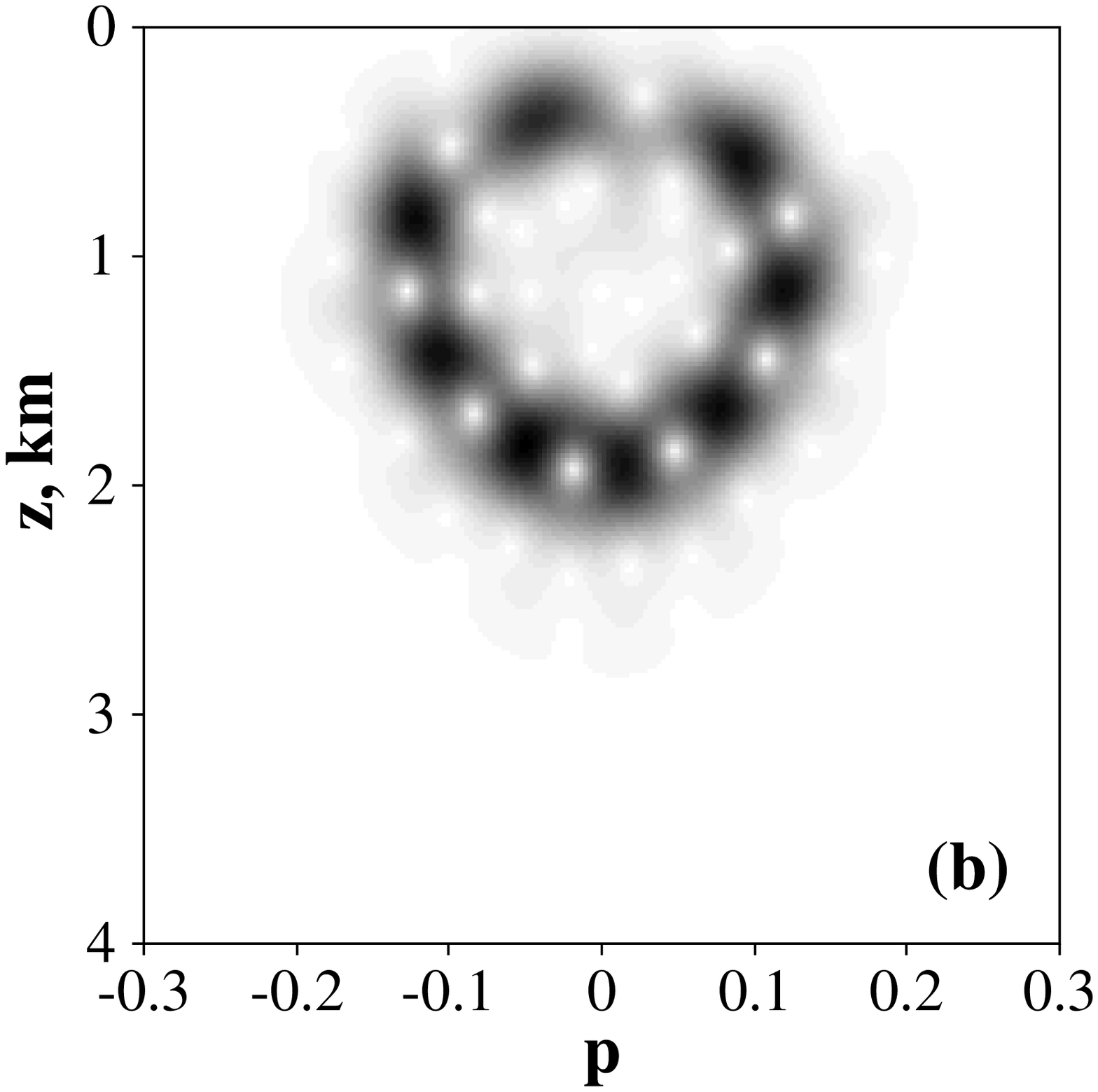}
\caption{Floquet modes with ``dial-plate'' structure. The signal frequency
is of 100 Hz.}
\label{eight100}
\end{center}
\end{figure}
Regular Floquet modes were found in all the phase space
corresponding to guided rays, even inside the chaotic sea.
Let us pay attention on those Floquet modes,
which have a ``dial-plate'' structure,
as that is shown in Fig.~\ref{eight100}.
This structure consists of eight well-resolved and ordered peaks.
The angular locations of the peaks in
Figs.~\ref{eight100}(a) and \ref{eight100}(b)
are different; the peaks in the upper plot correspond
to the zeroes in the lower one, and vice versa.

\begin{figure}[!t]
\begin{center}
\centerline{
\includegraphics[width=0.45\textwidth]{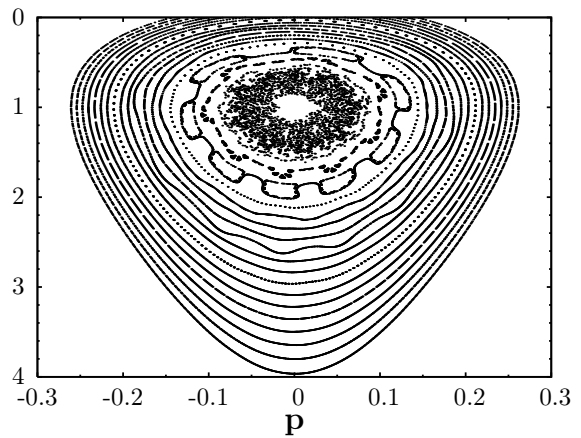}}
\caption{Poincar\'e map with $\varepsilon=0.0005$.}
\label{poinc2}
\end{center}
\end{figure}

\begin{figure}[!t]
\begin{center}
\includegraphics[width=0.45\textwidth]{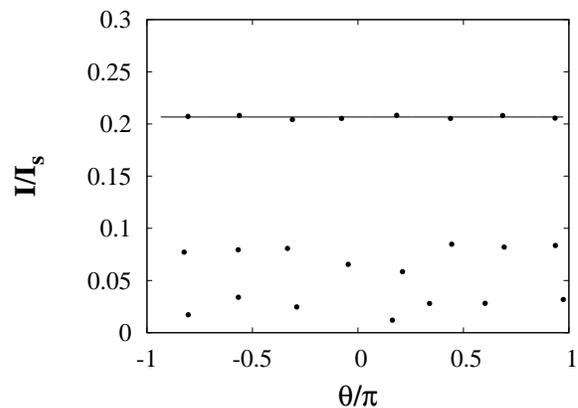}
\caption{Distribution of Husimi zeroes for the Floquet mode depicted in
Fig.~\ref{eight100}(a).
The horizontal line corresponds to the resonance $1:8$.
}
\label{zero2}
\end{center}
\end{figure}

The origin of the peaks can be clarified
if we construct the Poincar\'e map with decreased
amplitude of the range-dependent perturbation $\varepsilon=0.0005$.
This map is presented in Fig~\ref{poinc2}.
A bare comparison of Fig.~\ref{eight100} and Fig.~\ref{poinc2}
yields that peaks in Fig.~\ref{eight100}(a) are allocated
near the elliptic fixed points of the resonance $1:8$, while
peaks in Fig.~\ref{eight100}(b) are placed near the hyperbolic ones.
As it follows from Fig.~\ref{poinc},
this resonance is completely destroyed with $\varepsilon=0.005$,
and all its elliptic
periodic orbits of the resonance $1:8$ are unstable.
Another argument for the link between ``dial-plate'' structure
and resonance $1:8$ is presented in Fig.~\ref{zero2}.
There it is shown that eight Husimi zeroes of the mode, presented in
Fig.~\ref{eight100}(a), lay on the horizontal line $I=I_\text{res}$,
where $I_\text{res}$ is the resonant action, satisfying the equation
\begin{equation}
m\omega(I=I_\text{res})=nk_x,
\label{rescond}
\end{equation}
with $m=8$ and $n=1$. Here $\omega$ is the spatial frequency of ray oscillations
in the waveguide.
For the resonance $1:8$ we have $I_\text{res}\simeq0.2I_s$, where
$I_s$ is the most accessible action for rays propagating without
reflections from the lossy bottom.
Hence one may conclude that the peaks in the Husimi plots
relate to the so-called scarred states \cite{Heller,Bogomolny}.
\begin{figure}[!t]
\begin{center}
\includegraphics[width=0.45\textwidth]{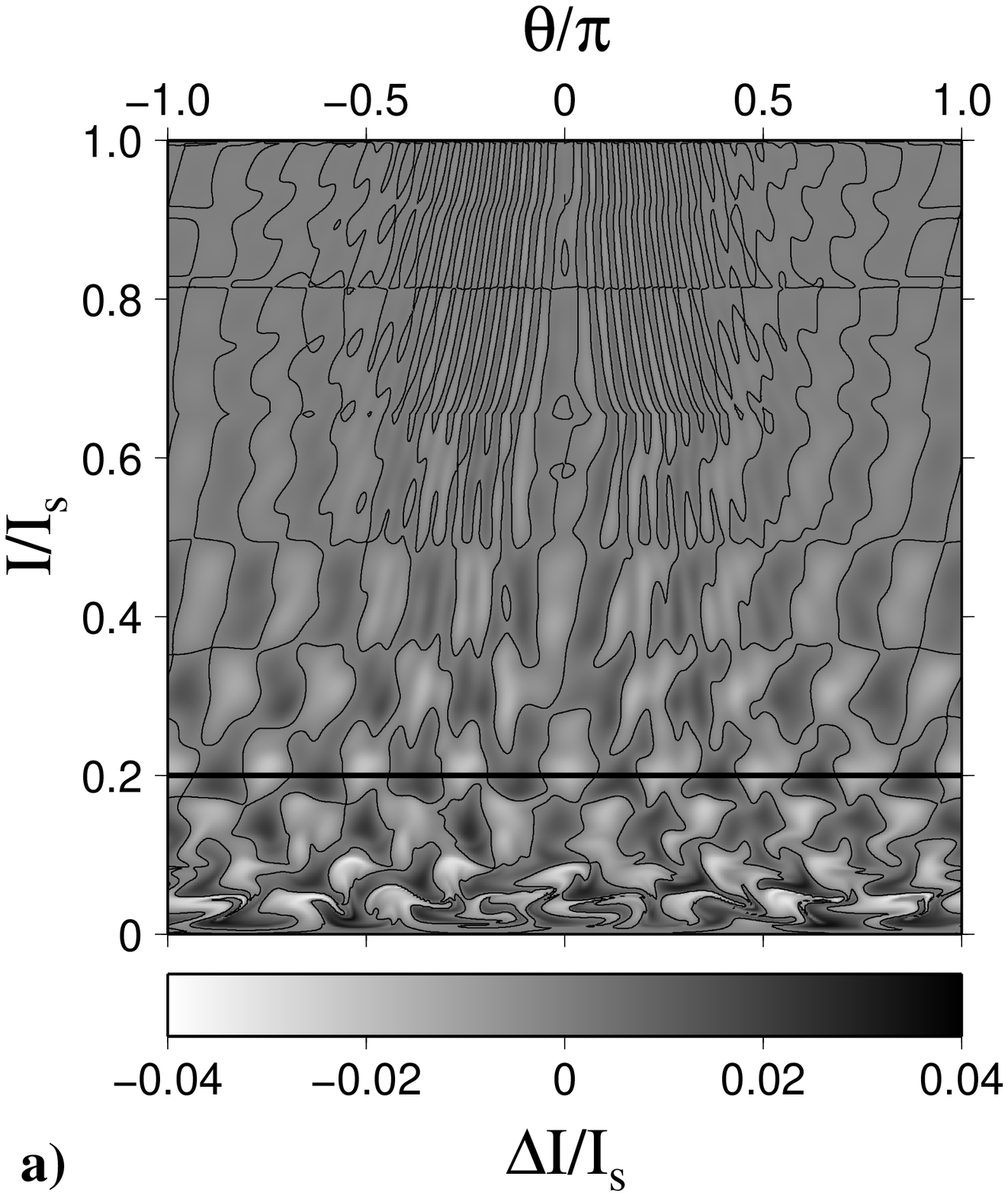}
\includegraphics[width=0.45\textwidth]{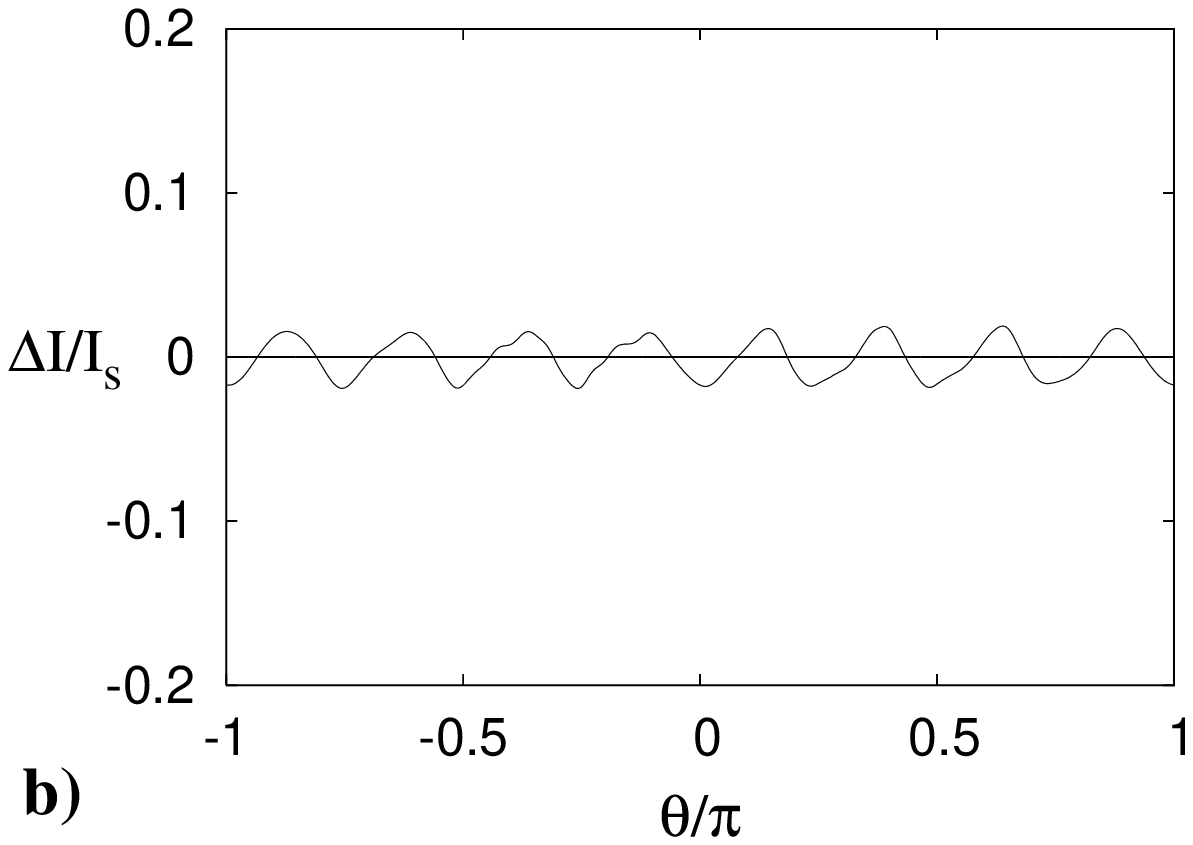}
\caption{(a) Variations of the normalized action
$\Delta I/I_s$ per ray cycle length in the plane
of normalized initial values of the action and angle with $\varepsilon=0.0005$.
Horizontal bold line corresponds to KAM resonance $1:8$.
Thin lines mark zero variations of the action.
(b) Variations of the normalized action along the
bold line.}
\label{acmap}
\end{center}
\end{figure}
\begin{figure}[!t]
\begin{center}
\includegraphics[width=0.45\textwidth]{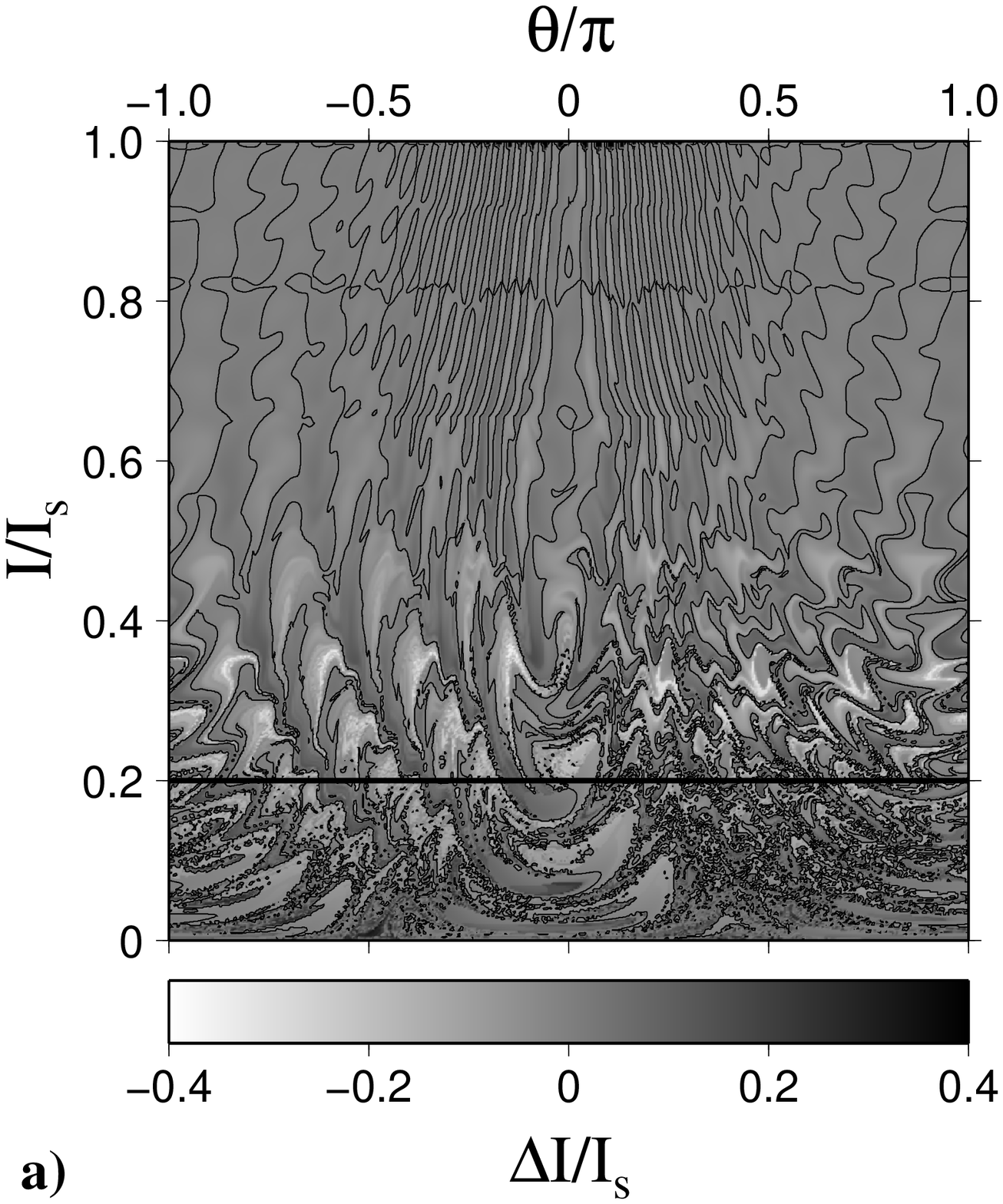}
\includegraphics[width=0.45\textwidth]{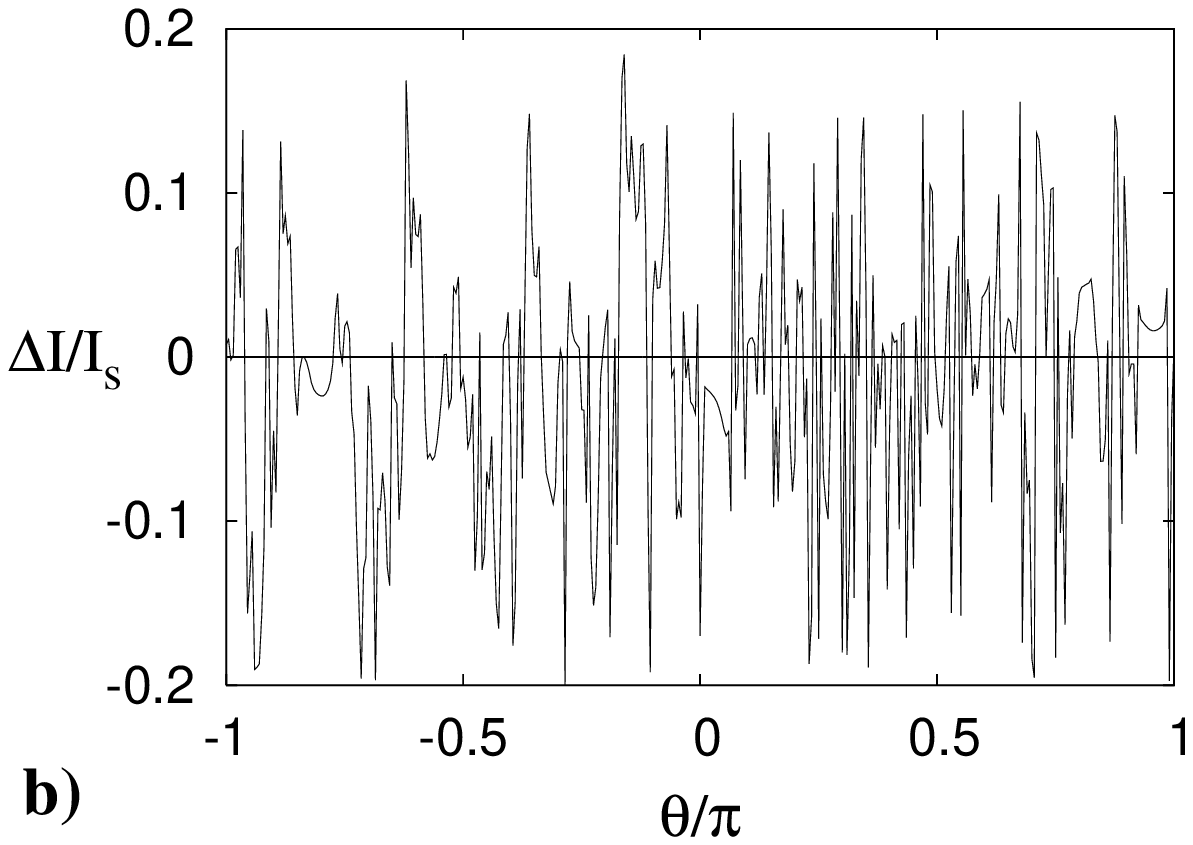}
\caption{The same as in Fig.~\ref{acmap}, but with $\varepsilon=0.005$.}
\label{acmap2}
\end{center}
\end{figure}

However this interpretation is problematic,
because increasing of $\varepsilon$ drastically alters phase space
distribution of periodic orbits.
We demonstrate it by constructing the map
representing variations of ray action per ray cycle length as function
of initial coordinates in phase space \cite{Chaos,JPA}.
These maps can be used for detecting periodic orbits.
The respective plots with $\varepsilon=0.0005$ and $\varepsilon=0.005$
are presented in Fig.~\ref{acmap} and Fig.~\ref{acmap2}.
All the periodic orbits belong to the zero lines
separating the regions of negative and positive variations
of action.
Approximate locations of elliptic and hyperbolic periodic orbits
may be found as
intersections of the zero lines with horizontal lines $I/I_s=I_\text{res}/I_s$.
Figures \ref{acmap}(b) and \ref{acmap2}(b)
represent variations of the action along that line.

\begin{figure}[!t]
\begin{center}
\includegraphics[width=0.45\textwidth]{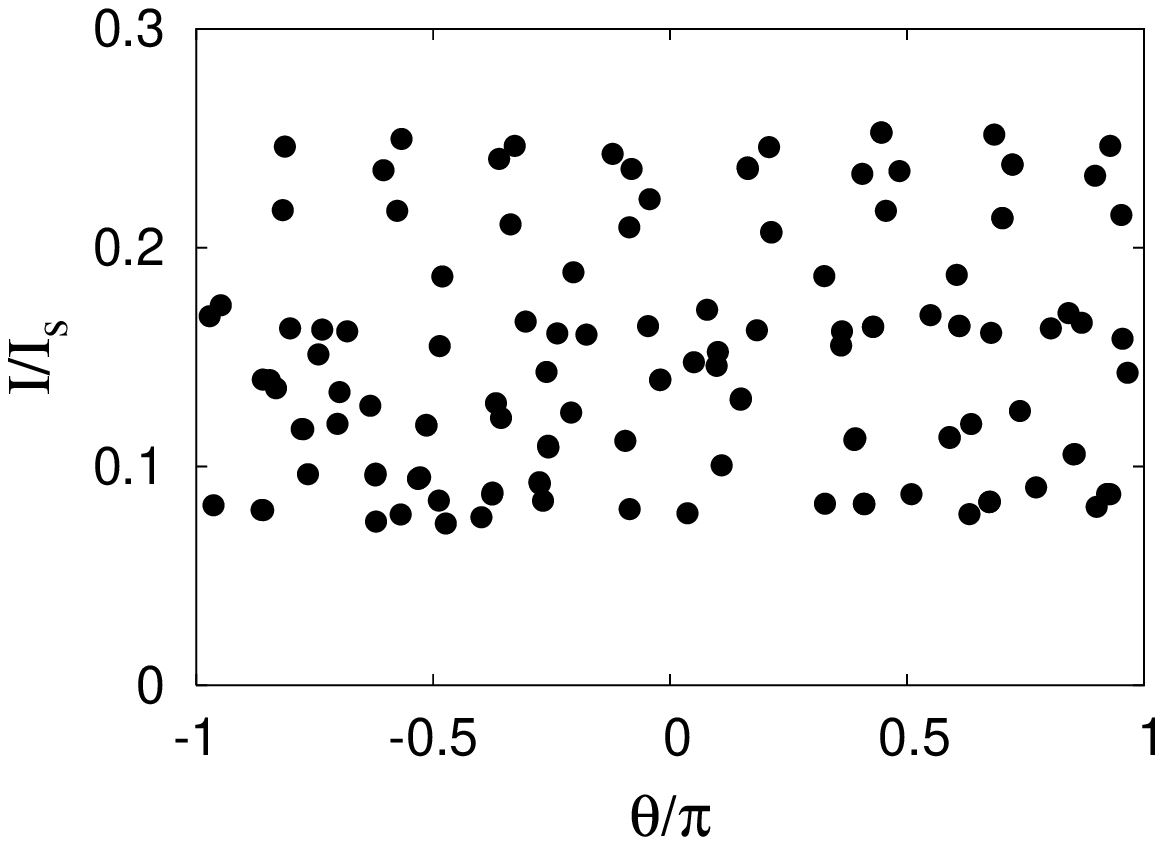}
\caption{Phase space locations of periodic orbits, computed with $\varepsilon=0.005$.}
\label{orbits}
\end{center}
\end{figure}

As it follows from Fig.~\ref{acmap}, the map is smooth
with $\varepsilon=0.0005$, and intersections with the horizontal bold line
$1:8$ are well ordered. In contrast,
the map with $\varepsilon=0.005$ (see Fig.~\ref{acmap2}) displays very complicated
``wave-like'' pattern, and intersections with the bold line
have dense and irregular distribution.
Obviously, these intersections cannot be associated
with elliptic and hyperbolic fixed points of KAM resonance $1:8$.
As a consequence, periodic orbits are disordered.
It is demonstrated in Fig.~\ref{orbits}, where we show
phase space locations of periodic orbits in the range of low values
of the action.
Evidently, the bright spots in the Husimi plots for the Floquet modes don't
correspond to certain periodic orbits.

\begin{figure}[!t]
\begin{center}
\includegraphics[width=0.45\textwidth]{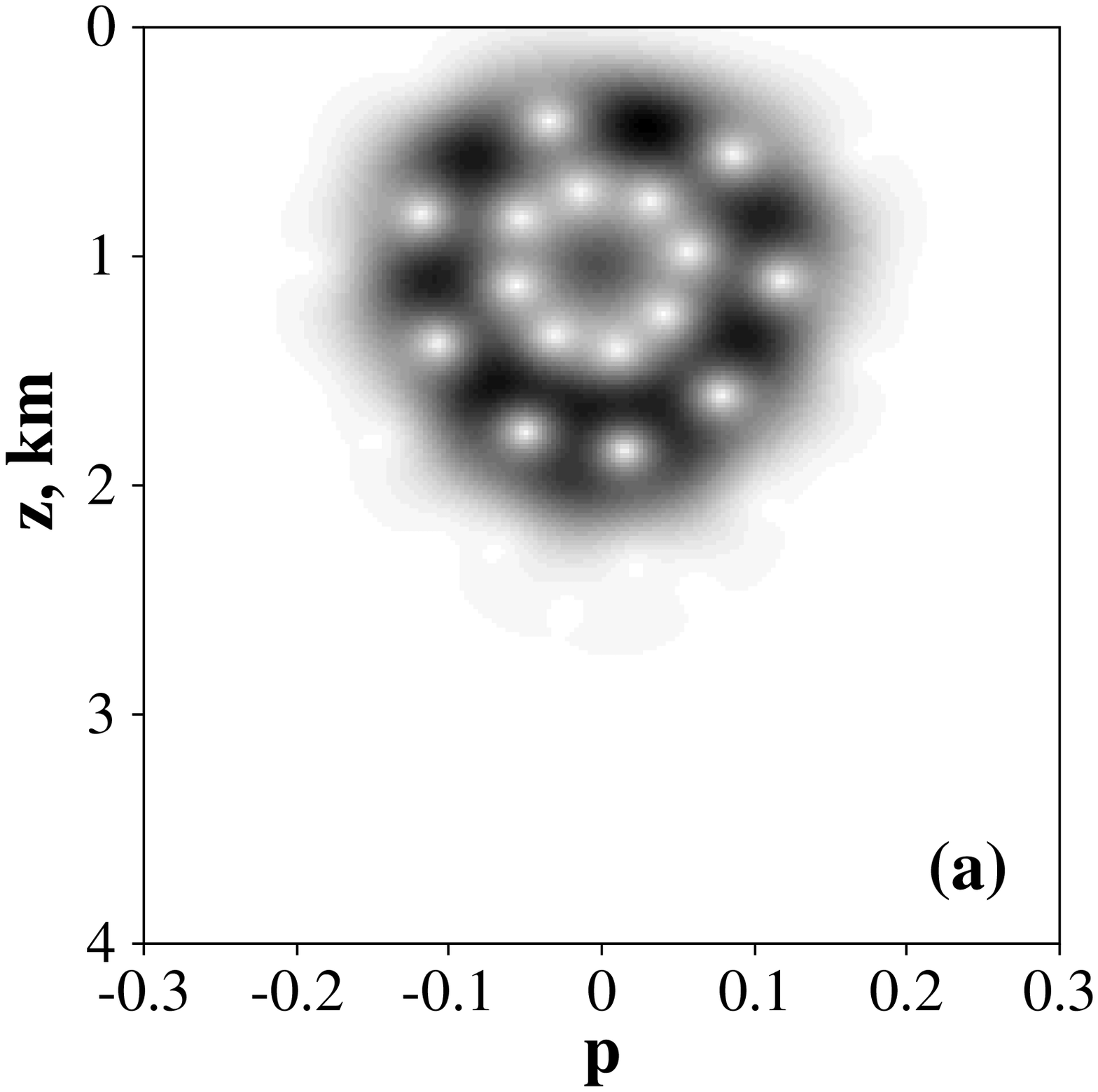}
\includegraphics[width=0.45\textwidth]{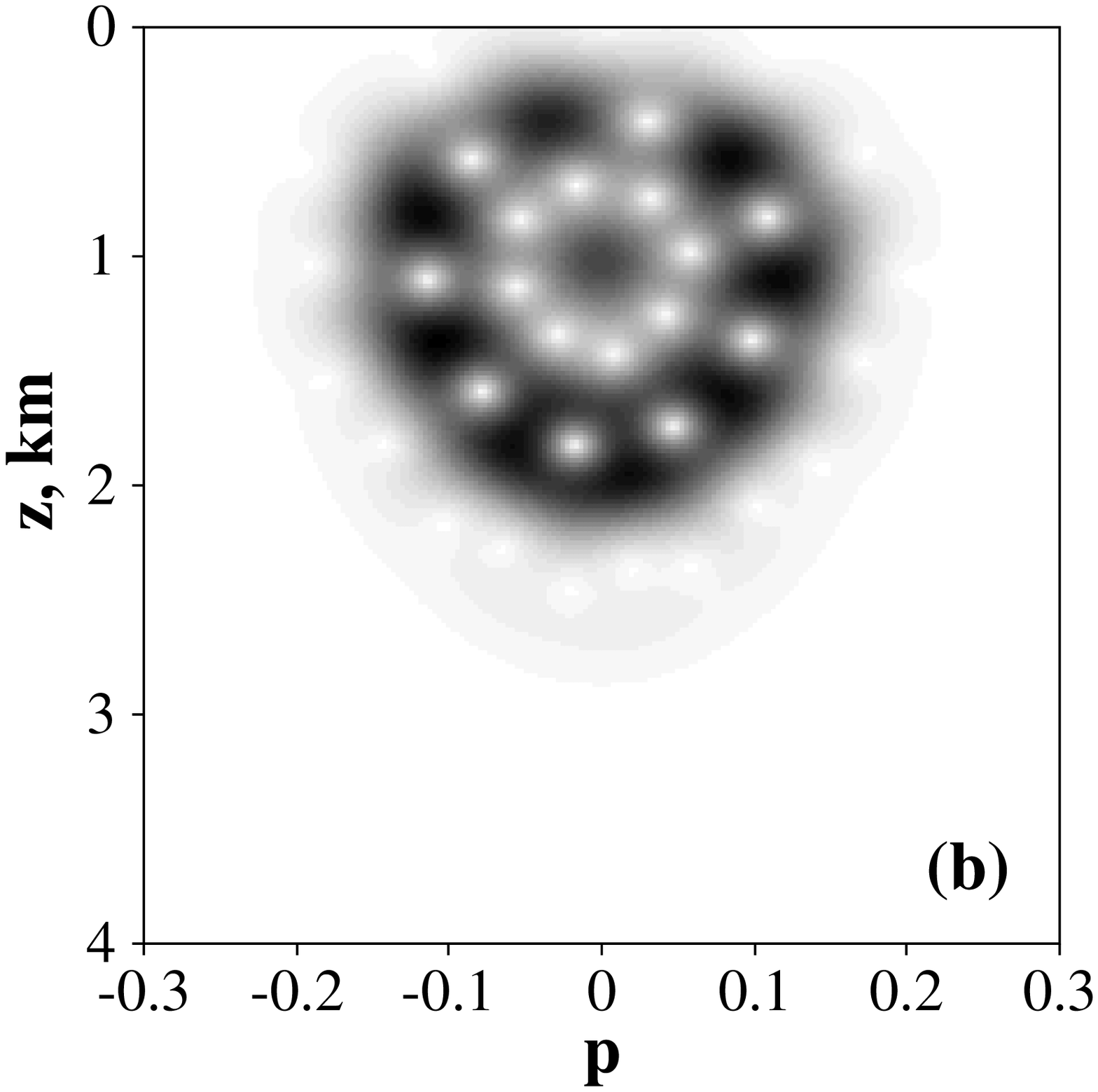}
\caption{Floquet modes with ``dial-plate'' structure. The signal frequency
is of 70 Hz.}
\label{eight70}
\end{center}
\end{figure}

\begin{figure}[!t]
\begin{center}
\includegraphics[width=0.45\textwidth]{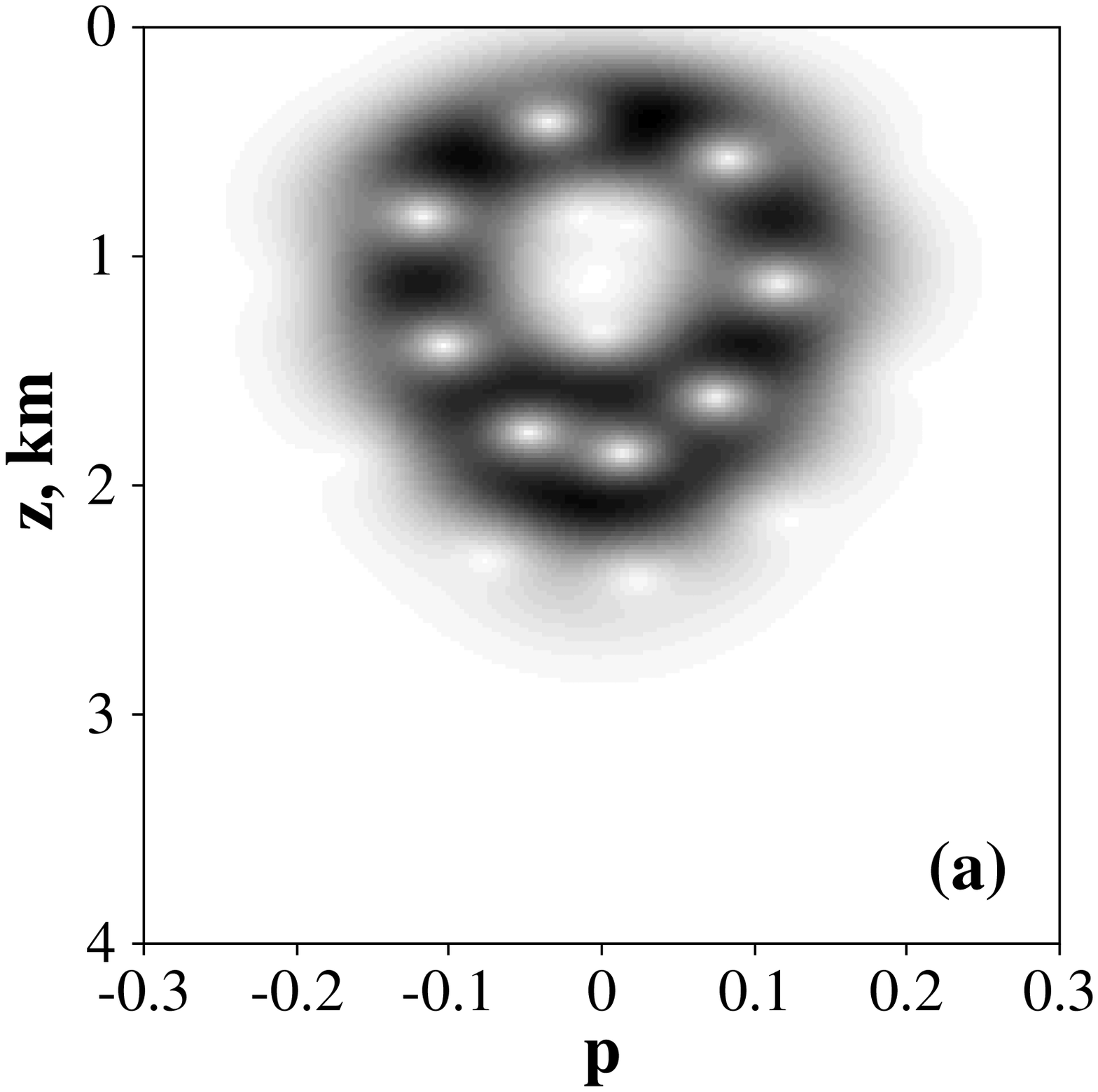}
\includegraphics[width=0.45\textwidth]{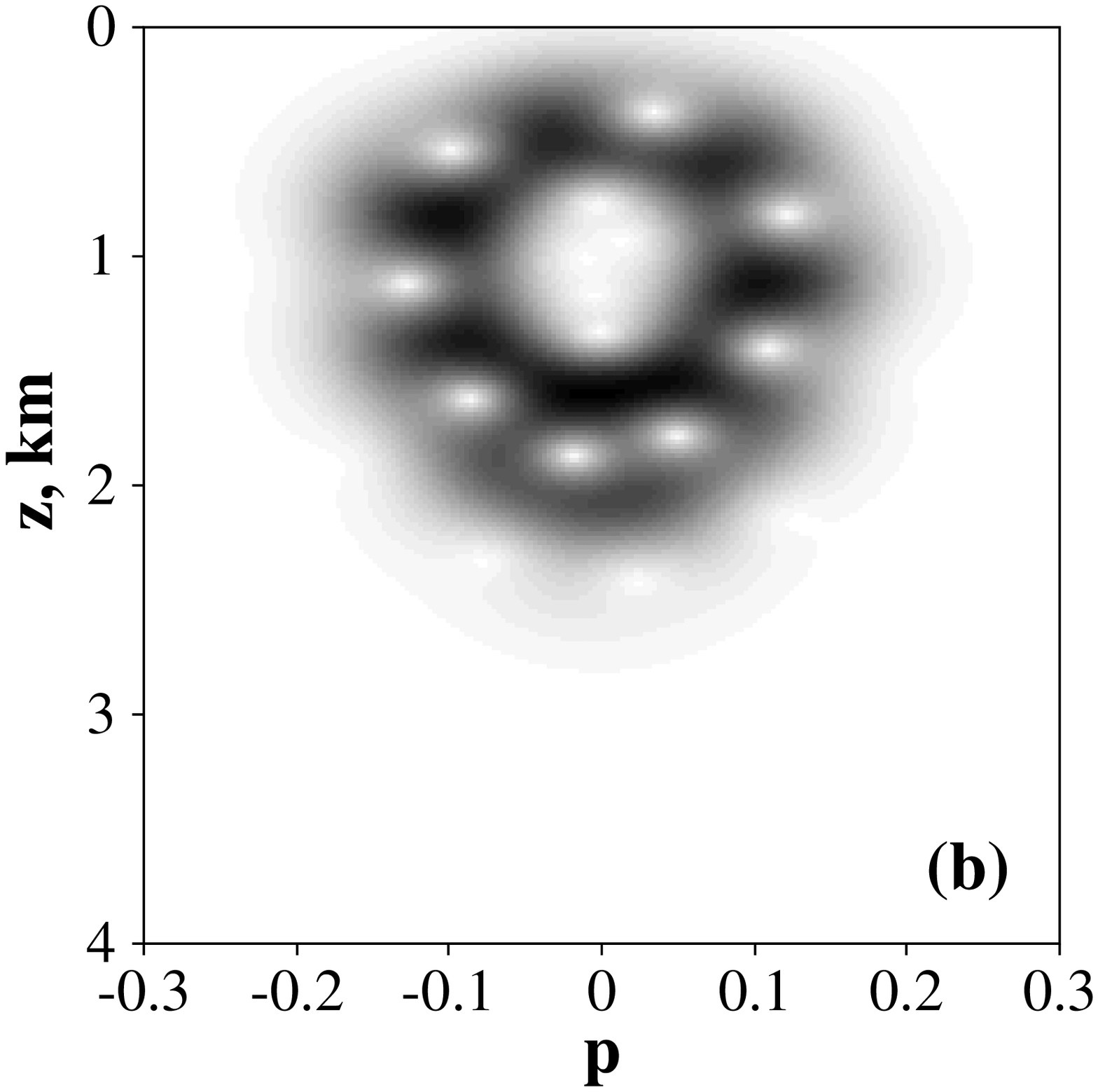}
\caption{Floquet modes with ``dial-plate'' structure.
The signal frequency is of 50 Hz.}
\label{eight50}
\end{center}
\end{figure}

Floquet modes having the same ``dial-plate'' structure
were also found with the frequencies of 70 and 50 Hz
(see Figs.~\ref{eight70} and \ref{eight50}).
In these cases 
Floquet modes don't possess any hallmarks of chaos.


Our explanation of the well-ordered peaks in the Husimi plots
is the following.
Decreasing of the signal frequency makes wave refraction
less sensitive to small scale vertical oscillations of the
sound-speed perturbation.
As it was shown in the previous section,
these oscillations are responsible for strong chaos
of near-axial rays.
Hence, the main source of chaos is suppressed at low frequencies.
Enhancing of ray-wave correspondence requires smoothing of fast vertical
oscillations using some averaging technique.
Evidently, the averaging of fast oscillations
means replacement of the original sound-speed perturbation
by the smoothed one with lower amplitude.
This implies that the disordered multiplication
of periodic orbits should be removed, and 
the well-ordered
periodic orbits of completely destroyed resonance $1:8$ should be
restored.
Thus the decreasing of frequency leads to the similar effect as
the decreasing of the perturbation's amplitude and is followed
by recovery of periodic orbits and occurrence
of the peaks observed. In some sense,
peaks of the Floquet modes may be regarded as some specific kind
of scars.

\section{Conclusion}

In the present paper we have considered sound wave motion
in an acoustic waveguide with the a range-dependent
sound-speed perturbation imposed.
It is shown that small-scale vertical oscillations
of the perturbation influence near-axial rays
in the resonant way.
Scattering on resonance makes near-axial rays unstable and leads
to the forming of a wide chaotic sea in the underlying phase space,
without any considerable islands of stability.
Moreover, small-scale vertical oscillations of the perturbation
cause proliferation and disordered phase space distribution of periodic orbits.
Nevertheless, the majority of the Floquet modes calculated
with frequencies of 50-100 Hz
reveals regular pattern with ordered peaks and Husimi zeroes.
The peaks are located near the elliptic or hyperbolic
fixed points of the ``classical'' KAM resonance $1:8$, that is
confirmed by constructing the Poincar\'e map with lower
amplitude of the perturbation.
However the peaks observed are not associated with certain periodic orbits
in the original case.
That resonance, as well as its periodic orbits, is completely 
destroyed with non-decreased amplitude of the sound-speed perturbation.
Thus we are faced with an eccentric conflict of ray and wave descriptions:
the peaks have specific ``classical'' resonant topology
but are not supported by classical periodic orbits.

In our opinion,
the revival of KAM resonance $1:8$ on the Husimi plots indicates
the necessity of frequency-dependent
corrections to the standart ray approximation, when ray motion
is strongly affected by small-scale features.
A likely way of introducing such corrections
is the construction of some effective sound-speed profile,
using the homogenization procedures \cite{Allaire},
or exploiting the quantum action \cite{Caron}.
We suppose that ray modeling with the effective sound-speed 
profile should be better consistent with wave pattern.
In particular, a corrected ray approximation may 
provide the desirable link between the Husimi peaks and 
the respective periodic orbits.
Our expectations are partially supported by
the numerical simulation, presented in \cite{Hege},
where it was shown that smoothing of fine-scale structures
doesn't lead to significant changes in a wavefield.
As a concluding remark,
it should be mentioned that 
the stabilizing of wave refraction with decreasing frequency
seems to be worth for overcoming limitations on the hydroacoustical
tomography, which are posed by ray chaos \cite{Tappert}.

\section*{Acknowledgments}

This work was supported by the projects of the President of the Russian Federation,
by the Program ``Mathematical Methods in
Nonlinear Dynamics'' of the Prezidium of the Russian Academy of Sciences,
and by the Program for Basic Research of the Far Eastern Division of
the Russian Academy of Sciences.
Authors are grateful to S.V.~Prants, A.I.~Neishtadt, A.L.~Virovlyansky
and A.I.~Gudimenko for helpful discussions
during the course of this research.

\section*{Appendix}

The action and angle
variables for the rays propagating without
reflections from the ocean surface are given by the formulae
\begin{equation}
I=\frac{b}{a}\left(\frac{\mu+\gamma}{2}-\sqrt{\mu\gamma-\frac{2E}{b^2}}
\right),
\label{action1}
\end{equation}
\begin{equation}
\theta=\pm
\frac{\pi}{2}\mp\theta_1,
\label{angle}
\end{equation}
where the quantities $Q$ and $\theta_1$ are given by the formulae
\begin{equation}
Q=\sqrt{(\mu-\gamma)^2+\frac{8E}{b^2}},
\label{Q}
\end{equation}
\begin{equation}
\theta_1=\arcsin\left(\frac{\mu+\gamma-(2\mu\gamma-4E/b^2)e^{az}}{Q}\right)
\label{vartheta2}.
\end{equation}
The upper and the lower signs in (\ref{angle}) correspond to positive
and negative values of ray momentum, respectively.
The action and the angle for surface-bounce rays are
given by the following formulae
\begin{equation}
\begin{aligned}
I=&\frac{b}{a}
\biggl(\biggr.\frac{\mu+\gamma}{4}-
\frac{\mu+\gamma}{2\pi}\arcsin{\dfrac{\mu+\gamma-2}{Q}}-\\
&-\frac{\pi+2\theta_2}{2\pi}
\sqrt{\mu\gamma-\frac{2E}{b^2}}
\biggl.\biggr)+\frac{|p(z=0)|}{\pi a},
\end{aligned}
\label{action2}
\end{equation}
\begin{equation}
\begin{aligned}
\theta=\pm\pi\frac{\theta_2-\theta_1}{\theta_2+\pi/2}.
\end{aligned}
\label{angl_ref}
\end{equation}
In (\ref{action2}), (\ref{angl_ref}) we used the notation
\begin{equation}
\theta_2=\arcsin\left(\frac{\mu+\gamma-2\mu\gamma+4E/b^2}{Q}\right)
\label{theta_r}.
\end{equation}
Under reflections, ray momentum
is given by the formula
\begin{equation}
p(z=0)=\pm\sqrt{2E-b^2(\mu-1)(\gamma-1)}.
\label{p0}
\end{equation}
The inverse transformation for the rays, propagating without reflections from the surface,
is expressed as follows
\begin{equation}
z(I,\,\theta)=\dfrac{1}{a}\ln
{\dfrac{a^2b^2\,\left(\mu+\gamma-Q\cos{\theta}\right)}{2\omega^2}},
\label{zI}
\end{equation}
\begin{equation}
p(I,\,\theta)=
\dfrac{\omega\,Q\sin{\theta}}
{a\left(\mu+\gamma-Q\cos{\theta}\right)},
\label{pI}
\end{equation}
where $\omega$ is the spatial frequency of ray oscillations in a waveguide.
It depends on $E$ in the following way:
\begin{equation}
\omega=ab\sqrt{\mu\gamma-2E/b^2}.
\label{wI}
\end{equation}


\begin{thebibliography}{100}

\bibitem{Wilc}
P.B.~Wilkinson, T.M.~Fromhold, R.P.~Taylor, and A.P.~Micolich,
Phys.~Rev.~E. {\bf 64}, 026203 (2001).
\bibitem{Klimes} L.~Klimes,
Pure and Appl. Geophys. {\bf 159}, 1465 (2002).
\bibitem{Bott} M.~Bottigliery, S.~De~Martino,
M.~Falanga, and C.~Godano,
Nonl.~Proc. Geophys. {\bf 12}, 1003 (2005).
\bibitem{Zas-ufn} S.S.~Abdullaev and G.M.~Zaslavsky,
Sov. Phys. Usp. {\bf 34}, 645 (1991).
\bibitem{Shtockman} H.J.~St\"ockman,
{\it Quantum Chaos: An Introduction}
(Cambridge University Press, Cambridge, England, 1999).
\bibitem{Chigarev} A.V.~Chigarev and Yu.V.~Chigarev,
Sov.Phys.Acoust. {\bf 24}, 432 (1978).
\bibitem{Palmer} D.R.~Palmer, M.G.~Brown, F.D.~Tappert, and H.F.~Bezdek,
Geophys.Res.Lett. {\bf 15}, 569 (1988).
\bibitem{Review}
M.G. Brown, J.A. Colosi, S.~Tomsovic, A.L.~Virovlyansky,
M.A.~Wolfson, and G.M.~Zaslavsky,
J.~Acoust.~Soc.~Am. {\bf 113}, 2533 (2003).
\bibitem{AET}
F.J.~Beron-Vera,
M.G. Brown, J.A. Colosi, S.~Tomsovic, A.L.~Virovlyansky,
M.A.~Wolfson, and G.M.~Zaslavsky,
J.~Acoust.~Soc.~Am. {\bf 114}, 1266 (2003).
\bibitem{Viro2001}
I.P.~Smirnov, A.L.~Virovlyansky, and G.M.~Zaslavsky,
Phys.~Rev.~E. {\bf 64}, 036221 (2001).
\bibitem{PRE} D.V. Makarov, M.Yu. Uleysky, M.V.~Budyansky, and S.V.~Prants,
Phys.~Rev.~E {\bf 73}, 066210 (2006).
\bibitem{JPA} D.V. Makarov and M.Yu. Uleysky, J.~Phys.~A: Math.~Gen. {\bf 39}, 489 (2006).
\bibitem{Chaos} D.V. Makarov, M.Yu. Uleysky, and
S.V.~Prants, Chaos {\bf 14}, 79 (2004).
\bibitem{ViroPRE99}
A.L.~Virovlyansky and G.M.~Zaslavsky,
Phys.~Rev.~E. {\bf 59}, 1656 (1999).
\bibitem{ViroPRE2005}
I.P.~Smirnov, A.L.~Virovlyansky, M.~Edelman, and G.M.~Zaslavsky,
Phys.~Rev.~E. {\bf 72}, 026206 (2005).
\bibitem{Viro-times} A.L.~Virovlyansky,
J.~Acoust.~Soc.~Am. {\bf 113}, 2523 (2003).
\bibitem{Brown}
F.J.~Beron-Vera,
M.G. Brown, J.~Acoust.~Soc.~Am. {\bf 115}, 1068 (2004).
\bibitem{Vadov} R.A.~Vadov,
Acoust. Phys. {\bf 46}, 544 (2000).
\bibitem{Acoust} D.V.~Makarov and M.Yu.~Uleysky,
Acoust. Phys. {\bf 53}, 495 (2007).
\bibitem{Latka} M. Latka, P. Grigolini, and B.J.~West,
Phys.~Rev.~A {\bf 47}, 4649 (1993).
\bibitem{BZ} B.~Sundaram and G.M.~Zaslavsky,
Chaos {\bf 9}, 483 (1999).
\bibitem{Backer}
A.~B\"acker, R.~Ketzmerick, and A.G.~Monastra,
Phys.~Rev.~Lett. {\bf 94}, 054102 (2005).
\bibitem{Hege}
K.C.~Hegewisch, N.R.~Cerruti, and S.~Tomsovic,
J.~Acoust.~Soc.~Am. {\bf 117}, 1582 (2005).
\bibitem{Itin} A.P.~Itin, A.I.~Neishtadt, and A.A.~Vasiliev,
Physica D. {\bf 141}, 281 (2000).
\bibitem{Neishtadt} A.I.~Neishtadt,
Proc. of Steklov Inst. of Math. {\bf 250}, 183 (2005).
\bibitem{JETPL} D.V.~Makarov and M.Yu.~Uleysky,
JETP Letters. {\bf 83}, 522 (2006).
\bibitem{Mezic} D.L.~Vainchtein, A.I.~Neishtadt, and I.~Mezic,
Chaos {\bf 16}, 043123 (2006).
\bibitem{PRE07} D.V.~Makarov and M.Yu.~Uleysky,
Phys.~Rev. E {\bf 75}, 065201(R) (2007).
\bibitem{CNS} D.V.~Makarov and M.Yu.~Uleysky,
in {\it Nonlinear Science and Complexity},
edited by A.C.~Luo, L.~Dai, H.R.~Hamidzadeh
(World Scientific Publishing Co, 2007);
Communications in Nonlinear Science and Numerical
Simulation {\bf 13}, 400 (2008).
\bibitem{Prants} S.V.~Prants and M.Yu.~Uleysky,
JETP Letters. {\bf 82}, 748 (2005).
\bibitem{ViroChaos2004}
I.P.~Smirnov, A.L.~Virovlyansky, and G.M.~Zaslavsky,
Chaos {\bf 14}, 317 (2004).
\bibitem{Ketz}
R.~Ketzmerick, L.~Hufnagel, F.~Steinbach,
and M.~Weiss,
Phys.~Rev.~Lett. {\bf 85}, 1214 (2000).
\bibitem{LebVor} P. Leboeuf and A.~Voros,
J. Phys.~A {\bf 23}, 1765 (1990).
\bibitem{Arranz} F.J. Arranz, F. Borondo, and R.M.~Benito,
Phys.~Rev.~E {\bf 54}, 2458 (1996).
\bibitem{Biswas} D.~Biswas and S.~Sinha,
Phys.~Rev.~E {\bf 60}, 408 (1999).
\bibitem{Sugita} A. Sugita and H.~Aiba,
Phys.~Rev.~E {\bf 65}, 036205 (2002).
\bibitem{Heller}
E.J.~Heller,
Phys.~Rev.~Lett. {\bf 53}, 1515 (1984).
\bibitem{Bogomolny}
E.B.~Bogomolny,
Physica D {\bf 31}, 169 (1988).
\bibitem{Allaire} G.~Allaire and M.~Vanninathan,
e-print arXiv:math-ph/0510083 (2005).
\bibitem{Caron} L.A. Caron, D. Huard,
G.~Melkonyan, K.J. Moriarty, and
L.P. Nadeau,
J.~Phys A.: Math. Gen. {\bf 37}, 6251 (2004).
\bibitem{Tappert}
F.D.~Tappert and Xin Tang,
J.~Acoust.~Soc.~Am. {\bf 99}, 185 (1996).

\end{thebibliography}
\end{document}